 \documentclass[aps,prl,twocolumn]{revtex4}
\usepackage{graphicx}
\usepackage{amsmath}
\usepackage{amssymb}
\usepackage{mathrsfs}
\usepackage{amsfonts}
\usepackage{dcolumn}
\usepackage{bm}
\newcommand{\ha}{\hat{a}}
\newcommand{\hb}{\hat{b}}

\providecommand{\abs}[1]{\lvert#1\rvert}

\providecommand{\abs}[1]{\lvert#1\rvert}

\begin{document}

\title{Tools for detecting entanglement between different degrees of freedom in quadrature squeezed cylindrically polarized modes}
\author{C. Gabriel$^{1,2}$, A. Aiello$^{1,2}$, S. Berg-Johansen$^{1,2}$, Ch. Marquardt$^{1,2}$ and G. Leuchs$^{1,2}$}

\affiliation{
$^{1}$ Max Planck Institute for the Science of Light, Guenther-Scharowsky-Str.\ 1, D-91058 Erlangen, Germany \\
$^{2}$ Institute of Optics, Information and Photonics, University Erlangen-Nuremberg, Staudtstr.\ 7/B2, D-91058 Erlangen, Germany \\
}
\begin{abstract}
Quadrature squeezed cylindrically polarized modes contain entanglement not only in the polarization and spatial electric field variables but also between these two degrees of freedom \cite{Gabriel2011}. In this paper we present tools to generate and detect this entanglement. Experimentally we demonstrate the generation of quadrature squeezing in cylindrically polarized modes by mode transforming a squeezed Gaussian mode. Specifically, $-1.2\,\mathrm{dB}\pm0.1\,\mathrm{dB}$ of amplitude squeezing are achieved in the radially and azimuthally polarized mode. Furthermore, theoretically it is shown how  the entanglement contained within these modes can be measured and how strong the quantum correlations, depending on the measurement scheme, are.  
\end{abstract} 
\maketitle
\section{Introduction}
\label{intro}
Nowadays continuous-variable entanglement can be generated in several different degrees of freedom (DOFs). For example, one has entangled the polarization \cite{Bowen2002,Dong2007}, spatial \cite{Boyer2008,Wagner2008} or quadrature \cite{Ou1992,Silberhorn2001} electric field variables respectively. However, continuous-variable entanglement between different DOFs, or so-called ``hybrid entanglement'', is a yet widely unexplored field. This is contrary to the field of  discrete-variables, where entanglement between DOFs has been subject to many investigations and  has been  successfully demonstrated experimentally \cite{zukowski1991,ma2009,neves2009,barreiro2010,nagali2010}. Recently, we have shown that one can generate continuous-variable hybrid entanglement by quadrature squeezing cylindrically polarized modes \cite{Gabriel2011}. These states extend the application spectrum of continuous-variable entangled systems and could be in particular useful for the generation of cluster states which are of great importance in quantum computing \cite{Raussendorf2003,Nielsen2004,Browne2005,Menicucci2006}. The entanglement between the different DOFs in quadrature squeezed cylindrically polarized modes is based on a classical ``structural inseparability'' of the polarization and spatial field variables  \cite{Gabriel2011}. In other words, already in a classical picture these modes cannot be described as a product state of the polarization and spatial DOF \cite{Holleczek2011,Qian2011}. 

\begin{figure*}
\center
\includegraphics[width=14cm]{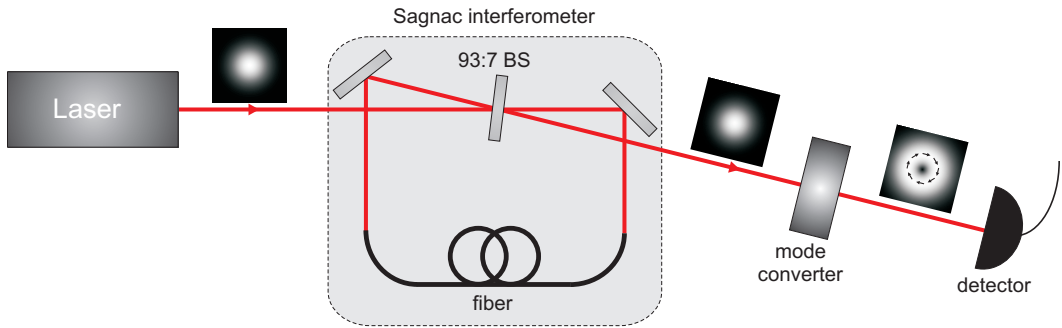}
\caption{The experimental scheme to generate amplitude squeezing in azimuthally and radially polarized modes. A femtosecond laser emits linearly polarized modes  with a Gaussian intensity profile at a wavelength of 1560\,nm. The light is injected into a Sagnac interferometer where the mode gets amplitude squeezed. Afterwards the Gaussian mode is converted into either a azimuthally or radially polarized mode. The amplitude quantum fluctuations of the generated modes are measured with a direct detection scheme [beam splitter (BS)].}
\label{Setup}       
\end{figure*}

In this paper we provide the tools necessary to characterize the hybrid entanglement present in quadrature squeezed cylindrically polarized modes. In the first part, we demonstrate the experimental realization of amplitude squeezing in radially and azimuthally polarized beams. These nonclassical states of light form the backbone for the generation of hybrid entanglement with cylindrically polarized modes.  In the second part, we present the measurement schemes to observe polarization, spatial and hybrid entanglement. Theoretically we carefully characterize the entanglement contained between the different DOFs and show how the  quantum correlations depend on the chosen measurement scheme.   

\section{Experimental realization of amplitude squeezed cylindrically polarized modes}
\label{sec:1}

A main problem of generating quadrature squeezing in cylindrically polarized modes, i.e. modes with a complex spatial as well as a complex polarization pattern,  is that many nonlinear media only interact with only one fixed polarization or   an inappropriate spatial mode. In \cite{Gabriel2011} we have shown that a specially tailored nanobore photonic crystal fiber is one of the few nonlinear media which can directly squeeze cylindrically polarized modes. Here we take a different approach. We first amplitude squeeze a linearly polarized mode with a Gaussian intensity profile and afterwards mode convert it into a cylindrically polarized mode of our choice. This scheme has the advantage that squeezing with Gaussian modes is a well established method as well as that the mode transformation allows a wider choice of cylindrically polarized modes to be generated as one is not fixed to the modes supported by the nonlinear medium.

Our experimental scheme is depicted in Fig.~\ref{Setup}. An asymmetric Sagnac interferometer \cite{Schmitt1998} generates an amplitude squeezed linearly polarized Gaussian mode. The Sagnac loop consists of a 93:7 beam splitter and a $6.45\,\mbox{m}\pm0.02\,\mbox{m}$ long  polarization maintaining   single-mode fiber (3M FS-PM-7811).  A shot-noise limited laser (ORIGAMI, Onefive GmbH) centered at a wavelength of $\lambda=1560$\,nm and emitting 220\,fs pulses acts as a light source. The quantum noise fluctuations of the light beam exiting the Sagnac interferometer are observed with a single detector  at a sideband frequency of $10.2$\,MHz. The coherent output beam of the laser is used to calibrate to the shot-noise level. With this method a total  of $-4.3\,\mathrm{dB}\pm0.1\,\mathrm{dB}$ of amplitude squeezing has been measured in the Gaussian mode. 

\begin{figure}
\center
\includegraphics[width=0.9\columnwidth]{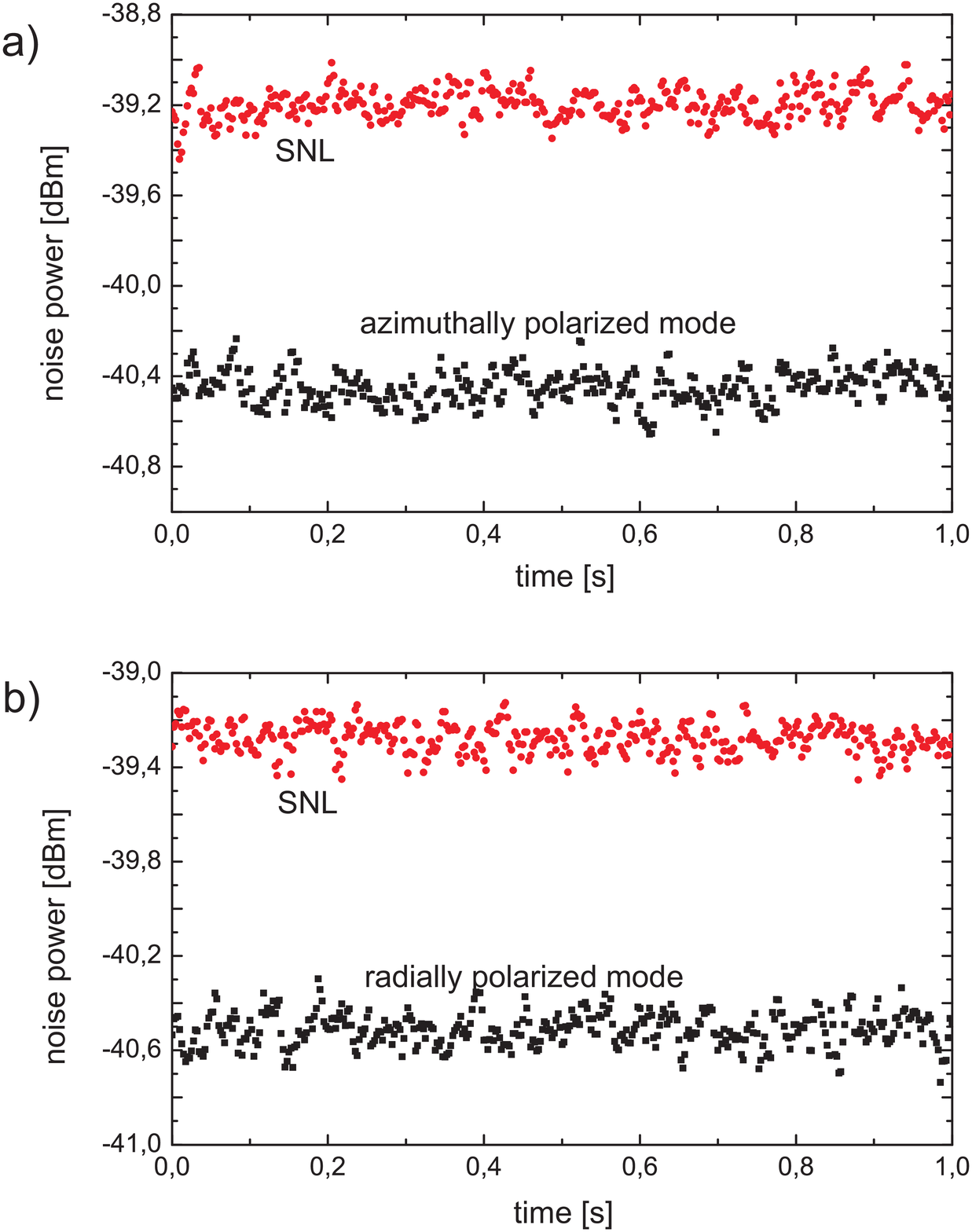}
\caption{The observed amplitude squeezing of a) the azimuthally and b) the radially polarized mode. The red data points represent the shot noise limit (SNL) while the black dots are the amplitude noise measurements of the corresponding cylindrically polarized mode.}
\label{sqdata}       
\end{figure}

In the next step, the amplitude squeezed Gaussian mode is sent through a liquid-crystal polarization converter (ARCoptix). This mode transformer is capable of converting linearly polarized Gaussian modes into  modes with either a continuous azimuthally or radially polarized polarization distribution. The generated modes are depicted in Fig.~\ref{modes}. After the mode converter $-1.2\,\mathrm{dB}\pm0.1\,\mathrm{dB}$ are observed in both the azimuthally and radially polarized mode. The results are shown in Fig~\ref{sqdata}. There are two main reasons why there is less amplitude squeezing contained in the cylindrically polarized modes than in the Gaussian mode: 1.~The mode transformation is not performed with unit efficiency but losses of about $~30\,\%$ occur during the conversion process. 2. The mode converter is not anti-reflection coated for 1560\,nm,  thus leading to losses of nearly $~30\,\%$ due to reflection. This means that in future the squeezing in cylindrically polarized modes can be significantly increased by improving the mode converter devices. 

The presented results show that the generation of efficient squeezing in different cylindrically polarized modes, namely the azimuthally and radially polarized modes, is feasible. This clearly increases the application spectrum of these modes in the wide field of quantum optics.

\section{Detecting entanglement between different degrees of freedom}
\label{sec:2}

Cylindrically polarized modes can be described with  the help of horizontally ($x$) or vertically ($y$) polarized first-order Hermite-Gauss modes (($10$) and ($01$)) associated to the annihilation operators $\ha_{x 01}$, $\ha_{y 01}$, $\ha_{x 10}$ and $\ha_{y 10}$.  The azimuthally and radially polarized modes are then given by the following annihilation operators  \cite{Holleczek2011}:
\begin{eqnarray}\label{Annih}
\ha_A &=&\frac{1}{2}\bigl( - \ha_{x 01} + \ha_{y 10}  \bigr), \\
\ha_R &=&\frac{1}{2}\bigl( \ha_{y 01} + \ha_{x 10}  \bigr).
\end{eqnarray}
In this section we quantify the entanglement generated inside these modes when they are quadrature squeezed. Furthermore, we present the experimental schemes with which one can detect the entanglement between the different DOFs. We will  consider in detail the exemplary case of a bright quadrature squeezed azimuthally polarized mode as this is one of the states we can generate experimentally. However, it should be noted that the  results here derived are valid for all cylindrically polarized modes and also apply to vacuum states.

A bright quadrature squeezed azimuthally polarized mode can be written as \cite{Gabriel2011}:
\begin{eqnarray}\label{Sq1}
|v_0, \zeta_0 \rangle_A =&& \hat{D}_{A}(v_0)\hat{S}_{A}(\zeta_0)| 0 \rangle \nonumber \\
=&& \hat{D}_{x01}(v)\hat{S}_{x01}(\zeta)\times \nonumber \\
&& \times \hat{D}_{y10}(v)\hat{S}_{y10}(\zeta) \hat{S}_{x01,y10}(-\zeta)| 0 \rangle,
\end{eqnarray}
where $\hat{S}_i(\zeta)=e^{(\zeta^* \hat{a}^2_i - \zeta \hat{a}^\dagger_i\/^2)/2}$ is the single-mode squeezing operator,   and $\hat{D}_{i}(v)= e^{v \ha_{i}^\dagger - v^* \ha_{i}}$ the displacement operator with $v_0=\sqrt{2} v$ being the displacement of the azimuthally polarized mode. The value $\zeta_0=2\zeta=re^{i\theta}$ quantifies the squeezing with $r$ being the degree of squeezing and $\theta$ the squeezing angle. The two-mode squeezing operator $\hat{S}_{x01,y10}(-\zeta)=e^{-(\zeta^* \ha_{y10} \hb_{x01}- \zeta  \hb_{x01}^\dagger \ha_{y10}^\dagger )}$ describes the entanglement existing between the orthogonally polarized Hermite-Gaussian modes.  In order to quantify this entanglement  an appropriate measure is required. A prominent one is the Simon-Duan inseparability criterion \cite{Duan2000,Simon2000} which can be extended to the Stokes operators  for a bipartite system $a\oplus b$ \cite{Gabriel2011}:
\begin{figure}
\center
\includegraphics[width=0.99\columnwidth]{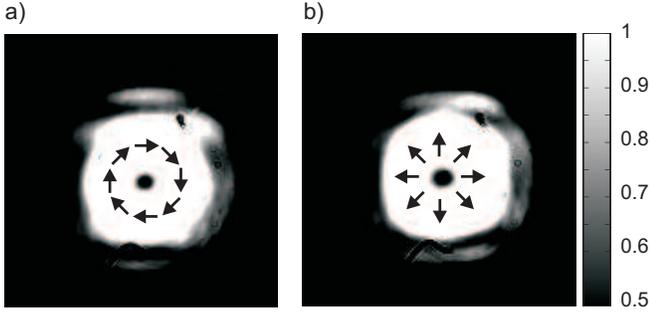}
\caption{The normalized intensity profiles of the generated a) azimuthally and b) radially polarized mode after the mode converter. The arrows indicate the local direction of the polarization.}
\label{modes}       
\end{figure}
\begin{eqnarray}
V(\hat{S}_{\sigma,DOF1}^{a}+\hat{S}_{\sigma,DOF2}^{b})+V(\hat{S}_{\rho,DOF1}^{a}-\hat{S}_{\rho,DOF2}^{b})<1, \nonumber \\
\label{uncrel} 
\end{eqnarray}
where $\hat{S}_{i}$ are the quantum Stokes operators corresponding to the classical Stokes parameters \cite{Luis2000,Hsu2009,Lassen2009}. Moreover, $(\sigma, \rho)=(1,2)$, $(1,3)$, or $(2,3)$ are the three combinations of the Stokes operators and $(DOF1,DOF2)=(pol,pol)$, $(spa,spa)$, $(pol,spa)$ and $(spa,pol)$ being the possible combinations of the spatial $(spa)$ or polarization $(pol)$ Stokes measurements on the two orthogonally polarized Hermite-Gaussian basis modes, one of which is in subsystem $a$ the other in subsystem $b$. Here $V(\hat{X})$ is the variance of the operator $\hat{X}$ normalized to $4\left| \alpha\right|$ with $\alpha= \mbox{cov}(\hat{S}^{a}_{\sigma},\hat{S}^{a}_{\rho})=\mbox{cov}(\hat{S}^{b}_{\sigma},\hat{S}^{b}_{\rho})$ being the covariance of the two Stokes operators. Depending on which violation occurs,  either  polarization, spatial  or hybrid entanglement  exists. 

In the following, we will theoretically derive how  polarization entanglement can be observed. The modifications needed in order to observe spatial and hybrid entanglement are minor and are presented at the end of this section.

\begin{figure*}
\center
\includegraphics[width=12cm]{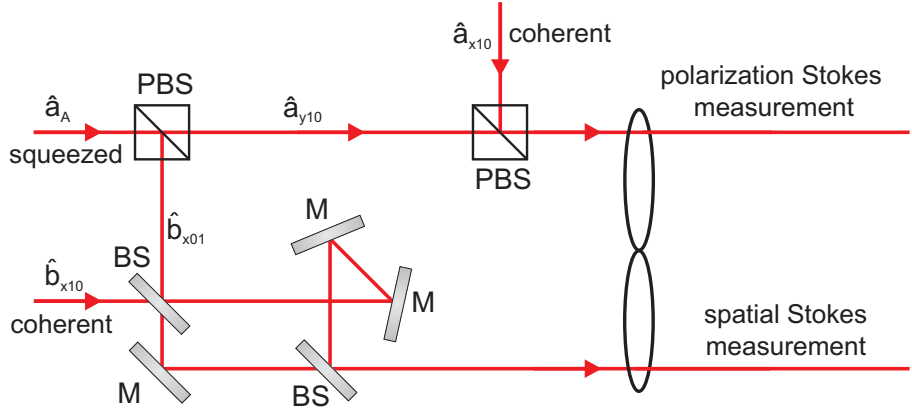}
\caption {Scheme to observe hybrid entanglement. In arm $a$ a polarization and in arm $b$ a spatial Stokes measurement is performed [polarizing beam splitter (PBS), mirror (M), 50:50 beam splitter (BS)].}
\label{hybridmeas}
\end{figure*}

To observe the quantum correlations contained within the azimuthally polarized mode, the two orthogonally polarized Hermite-Gaussian modes of the quadrature squee-zed azimuthally polarized beam need to be spatially separated into two arms $a$ and $b$. This can, for example, be performed with a polarizing beam splitter (PBS) as illustrated in Fig.~\ref{hybridmeas}. The state is then given by: 
\begin{eqnarray}\label{state2}
\lvert {v} ,{\zeta}\rangle_{a_{y10}b_{x01}} =  &&\hat{D}_{a_{y10}}(v) \hat{S}_{a_{y10}}(\zeta) 
\hat{D}_{b_{x01}}(-v)\times \nonumber \\ && \times \hat{S}_{b_{x01}}(\zeta) \hat{S}_{a_{y10} b_{x01}}(-{\zeta})\lvert 0 \rangle, 
\end{eqnarray} 
In order to access the polarization entanglement, each mode,  $\ha_{y10}$ and $\hb_{x01}$, is superimposed with a coherent beam with the same spatial mode but orthogonal polarization, namely $\ha_{x10}$ and $\hb_{y01}$. These have the amplitudes $w_1$ and $w_2$ respectively. Experimentally this can be performed by placing a PBS in each arm and inserting the coherent states in the second input ports of these as shown in Fig.~\ref{hybridmeas}. One can insert a half-wave plate orientated at $45 ^\circ$ in arm $b$  in order to achieve a polarization-symmetric state:
\begin{eqnarray}
&&\lvert {v} ,{\zeta}\rangle_{a_{y10}b_{y01}}\lvert {-w_1} \rangle_{a_{x10}}\lvert {w_2} \rangle_{b_{x01}} = \nonumber \\
&&=\hat{D}_{a_{x10}}(-w_1)\hat{D}_{a_{y10}}(v) \hat{S}_{a_{y10}}(\zeta)\hat{D}_{b_{x01}}(w_2)\times  \nonumber \\ && \times
\hat{D}_{b_{y01}}(-v) \hat{S}_{b_{y01}}(\zeta) \hat{S}_{a_{y10} b_{y01}}(-{\zeta})\lvert 0 \rangle.
\end{eqnarray}
The desired polarization Stokes parameter sets one wants to measure are \cite{korolkova2002,korolkova2005}:
\begin{subequations}\label{StokesA}
\begin{align}
\hat{S}_{0,pol}^a & = \ha_{x10}^\dagger \ha_{x10} + \ha_{y10}^\dagger \ha_{y10},\\  
\hat{S}_{1,pol}^a & = \ha_{x10}^\dagger \ha_{x10} - \ha_{y10}^\dagger \ha_{y10}, \\ 
\hat{S}_{2,pol}^a & = \ha_{x10}^\dagger \ha_{y10} + \ha_{y10}^\dagger \ha_{x10}, \\  
\hat{S}_{3,pol}^a & = \frac{1}{i} \bigl( \ha_{x10}^\dagger \ha_{y10} - \ha_{y10}^\dagger \ha_{x10} \bigr),
\end{align}
\end{subequations}
and
\begin{subequations}\label{StokesB}
\begin{align}
\hat{S}_{0,pol}^b & =  \hb_{x01}^\dagger \hb_{x01} + \hb_{y01}^\dagger \hb_{y01}, \\
\hat{S}_{1,pol}^b & = \hb_{x01}^\dagger \hb_{x01} - \hb_{y01}^\dagger \hb_{y01},\\
\hat{S}_{2,pol}^b & = \hb_{x01}^\dagger \hb_{y01} + \hb_{y01}^\dagger \hb_{x01}, \\
\hat{S}_{3,pol}^b & =\frac{1}{i} \bigl( \hb_{x01}^\dagger \hb_{y01} - \hb_{y01}^\dagger \hb_{x01} \bigr). 
\end{align}
\end{subequations}
From these one can determine the following quantities for the mean value of the Stokes operators:
\begin{subequations}\label{StokesAveA}
\begin{align}
\langle \hat{S}_0^a \rangle & = \abs{w_1}^2 + \frac{\abs{v_0}^2}{2} + 2 \mu \nu, \\
\langle \hat{S}_1^a \rangle  & = \abs{w_1}^2 - \frac{\abs{v_0}^2}{2} - 2 \mu \nu, \\
\langle \hat{S}_2^a \rangle  & = -\sqrt{2} \abs{w_1}\abs{v_0} \cos(\arg w_1 - \arg v_0),  \\
\langle \hat{S}_3^a \rangle  & = \sqrt{2} \abs{w_1}\abs{v_0} \sin(\arg w_1 - \arg v_0), 
\end{align}
\end{subequations}
and
\begin{subequations}\label{StokesAveB}
\begin{align}
\langle \hat{S}_0^b \rangle & = \abs{w_2}^2 + \frac{\abs{v_0}^2}{2} + 2 \mu \nu, \\
\langle \hat{S}_1^b \rangle  & = \abs{w_2}^2 - \frac{\abs{v_0}^2}{2} - 2 \mu \nu, \\
\langle \hat{S}_2^b \rangle  & = -\sqrt{2} \abs{w_2}\abs{v_0} \cos(\arg w_2 - \arg v_0),  \\
\langle \hat{S}_3^b \rangle  & = \sqrt{2} \abs{w_2}\abs{v_0} \sin(\arg w_2 - \arg v_0), 
\end{align}
\end{subequations}
 with $\mu=\mbox{cosh}\left(\frac{r}{2}\right)$ and $\nu=\mbox{sinh}\left(\frac{r}{2}\right)e^{i\theta}$. By, firstly,  defining the amplitude of the non-classical modes as  $v_0=\sqrt{n} \geq 0$, secondly, making the two coherent beams equally bright, i.e. $w_1 = w_2 = \sqrt{\frac{m}{2}} \geq 0$, and thirdly,  setting the squeezing angle to zero $\theta = 0$, the variances for the different Stokes parameters can be calculated:
\begin{eqnarray}
&& V(\hat{S}_{0,pol}^{a,b}) =   V_(\hat{S}_{1,pol}^{a,b})= \frac{1}{16} [ 4 n + 8 m -3 + \nonumber \\
&&+ 2(1 + 2 n) \cosh(2 r)+ \cosh(4 r)  - 4 n \sinh(2 r)], \\
&& V_(\hat{S}_{2,pol}^{a,b})= \frac{1}{4} [ 2 n +  m + \nonumber \\
&&+ (1 + m) \cosh(2 r)-1 - m \sinh(2 r) ], \\
&& V_(\hat{S}_{3,pol}^{a,b})= \frac{1}{4} [ 2 n +  m + \nonumber \\
&&+ (1 + m) \cosh(2 r)-1 + m \sinh(2 r) ]. 
\end{eqnarray}
The evaluation of the uncertainty relation, given in Eq.\,(\ref{uncrel}), reveals the  entanglement present between the different Stokes parameters. The following violations can be shown:
\begin{eqnarray}
\lim_{m \rightarrow \infty} V(\hat{S}_{2,pol}^{a}+\hat{S}_{2,pol}^{b})&+&V(\hat{S}_{3,pol}^{a}-\hat{S}_{3,pol}^{b}) =\nonumber \\
&=& \;e^{-r}\cosh r<1,
\label{violationpol1}
\end{eqnarray} 
and
\begin{eqnarray}
&& \lim_{m \rightarrow n} V(\hat{S}_{1,pol}^{a}+\hat{S}_{1,pol}^{b})+V(\hat{S}_{3,pol}^{a}-\hat{S}_{3,pol}^{b}) = \nonumber \\
&&=  \frac{1}{16 n} [ 12 n + 2(1 + 2 n) \cosh(2 r) + \cosh(4 r) - \nonumber \\
&&- 3 - 4 n \sinh(2 r)]<1, 
\label{violationpol2}
\end{eqnarray} 
depending on whether the coherent beams are much brighter than or have the same intensity as the two nonclassical beams. This proves that polarization entanglement between the two arms, $\ha_{y10}$ and $\hb_{x01}$, is present. The amplitude of the coherent beam determines between which Stokes parameters entanglement exists and how strong it is. 

Spatial entanglement can be shown in a very similar manner. In order to observe this, the two modes, $\ha_{y10}$ and $\hb_{x01}$, from the squeezed azimuthally polarized mode  have to be interfered with two coherent beams with the same polarization but orthogonal first-order Hermite-Gaussian modes, namely $\ha_{y01}$ and $\hb_{x10}$. This can be done with the help of an asymmetric Mach-Zehnder interferometer \cite{Hsu2009,Holleczek2011}. The Stokes operators for the spatial field variables are \cite{Padgett1999,Hsu2009,Lassen2009}
\begin{subequations}\label{StokesC}
\begin{align}
\hat{S}_{0,spa}^a & = \ha_{x10}^\dagger \ha_{x10} + \ha_{x01}^\dagger \ha_{x01},  \\
\hat{S}_{1,spa}^a & = \ha_{x10}^\dagger \ha_{x10} - \ha_{x01}^\dagger \ha_{x01}, \\
\hat{S}_{2,spa}^a & = \ha_{x10}^\dagger \ha_{x01} + \ha_{x01}^\dagger \ha_{x10},  \\
\hat{S}_{3,spa}^a & = \frac{1}{i} \bigl( \ha_{x10}^\dagger \ha_{x01} - \ha_{x01}^\dagger \ha_{x10} \bigr),
\end{align}
\end{subequations}
and
\begin{subequations}\label{StokesD}
\begin{align}
\hat{S}_{0,spa}^b & =  \hb_{y10}^\dagger \hb_{y10} + \hb_{y01}^\dagger \hb_{y01}, \\
\hat{S}_{1,spa}^b & = \hb_{y10}^\dagger \hb_{y10} - \hb_{y01}^\dagger \hb_{y01},\\
\hat{S}_{2,spa}^b & = \hb_{y10}^\dagger \hb_{y01} + \hb_{y01}^\dagger \hb_{y10}, \\
\hat{S}_{3,spa}^b & =\frac{1}{i} \bigl( \hb_{y10}^\dagger \hb_{y01} - \hb_{y01}^\dagger \hb_{y10} \bigr).
\end{align}
\end{subequations}

The calculations for the violations of the entanglement criterion  for the spatial Stokes operators are therefore very similar to the ones for the polarization Stokes parameters and yield the same violations of the inseparability criterion. With this knowledge it is straightforward to see that the violations shown in Eq.\,(\ref{violationpol1}) and Eq.\,(\ref{violationpol2}) of the Duan-Simon criterion are valid for any combination of the Stokes operators  made on each arm, thus proving that hybrid entanglement is also present.

A scheme to observe hybrid entanglement is illustrated in Fig.~\ref{hybridmeas}. In arm $a$ a polarization Stokes measurement is performed. For this purpose, the nonclassical mode $\ha_{y10}$ is interfered on a PBS with a coherent  beam $\ha_{x10}$. In arm $b$ a spatial Stoke measurement is performed by combining the nonclassical mode $\hb_{x01}$ with the coherent mode $\hb_{x10}$ in an asymmetric Mach-Zehnder interferometer. With the appropriate correlation measurements between the Stokes parameter pairs hybrid entanglement can then be observed. 

\section{Conclusion}
\label{sec:3}

In this paper we have provided the necessary tools to observe hybrid entanglement in quadrature squeezed cylindrically polarized modes. We have demonstrated that the mode transformation of an amplitude squeezed Gaussian mode with the help of a polarization converter is an efficient method to   generate amplitude squeezing in radially and azimuthally polarized modes. Furthermore, we have presented schemes to measure the polarization, spatial and hybrid entanglement contained within these modes. It depends on the measurement scheme between which Stokes parameters entanglement exists and how strong the correlations are. A great advantage of generating entanglement with quadrature squeezed cylindrically polarized modes is that one can always  freely choose which entanglement one would like to observe. The gained insights allow a more detailed understanding of the intriguing  quantum properties contained in cylindrically polarized modes. The results show that the realization of such states is indeed feasible and their implementation into  quantum information protocols absolutely plausible.

\end{document}